\documentclass[usenatbib]{mnras}
\usepackage[utf8]{inputenc}
\usepackage{aas_macros}
\usepackage{amsfonts}
\usepackage{amsmath}
\usepackage{bm}
\usepackage{amssymb}
\usepackage{graphicx}
\usepackage[dvipsnames]{xcolor}
\usepackage{xfrac}
\usepackage{hyperref}
\usepackage{nameref}
\usepackage{cleveref}
\usepackage{multirow}
\usepackage{soul}
\usepackage{url}
\crefformat{section}{\S#2#1#3}
\hypersetup{colorlinks=true,allcolors=teal}

\newcommand{\Mpch}{\ensuremath{h^{-1}{\rm Mpc}}}

\newcommand{\be}{\begin{equation}}
\newcommand{\ee}{\end{equation}}

\title[Lognormal simulations of Lyman-$\alpha$ forest]{Lognormal semi-numerical simulations of the Lyman-$\alpha$ forest: comparison with full hydrodynamic simulations} 
\author[Arya et al.]{
Bhaskar Arya$^{1}$\thanks{E-mail: bharya@iucaa.in}, 
Tirthankar Roy Choudhury$^{2}$\thanks{E-mail: tirth@ncra.tifr.res.in}, 
 Aseem Paranjape$^{1}$\thanks{E-mail: aseem@iucaa.in}, \&
 Prakash Gaikwad$^{3}$\thanks{E-mail: gaikwad@mpia-hd.mpg.de} 
 \\  
 $^1$ Inter-University Centre for Astronomy \& Astrophysics, Ganeshkhind, Post Bag 4, Pune 411007, India\\
 $^2$ National Centre for Radio Astrophysics, TIFR, Post Bag 3, Ganeshkhind, Pune 411007, India\\
 $^3$ Max-Planck-Institut für Astronomie, Königstuhl 17, D-69117 Heidelberg, Germany
 }

\begin{document}
\label{firstpage}
\pagerange{\pageref{firstpage}--\pageref{lastpage}}
\maketitle

\begin{abstract}
\noindent

Observations of the Lyman-$\alpha$ (Ly$\alpha$) forest in spectra of distant quasars enable us to probe the matter power spectrum at relatively small scales. With several upcoming surveys, it is expected that there will be a many-fold increase in the quantity and quality of data, and hence it is important to develop efficient simulations to forward model these data sets. One such semi-numerical method is based on the assumption that the baryonic densities in the intergalactic medium (IGM) follow a lognormal distribution. In this work, we test the robustness of the lognormal model of the Ly$\alpha$ forest  in recovering a set of IGM parameters by comparing with high-resolution Sherwood SPH simulations. We study the recovery of the parameters $T_0$ (temperature of the mean-density IGM), $\gamma$ (slope of the temperature-density relation) and $\Gamma_{12}$ (hydrogen photoionization rate) at $z \sim 2.5$ using a Markov Chain Monte Carlo (MCMC) technique for parameter estimation. Using three flux statistics, the probability distribution, the mean flux and the power spectrum, values of all three parameters, $T_0$, $\gamma$ and $\Gamma_{12}$ implied in the SPH simulations are recovered within $1 - \sigma$ ($\sim$ 9, 4 and 1\% respectively) of the median (best-fit) values. We verify the validity of our results at different baryon smoothing filter, SNR, box size \& resolution, and data seed and confirm that the lognormal model can be used as an efficient tool for modelling the Ly$\alpha$ transmitted flux at $z \sim 2.5$.

\end{abstract}


\section{Introduction}
\label{sec:intro}
\noindent

The series of Ly$\alpha$ absorption seen towards distant quasars or galaxies is one of the most sensitive tools to probe the cosmological matter distribution in the Universe. The Ly$\alpha$ forest is sensitive to the baryon distribution at mildly non-linear densities $\Delta \lesssim 10$ \citep{viel+02, mattarrese_2002, demi_2011}, different from galaxies which probe highly non-linear regimes of the density. The observed Ly$\alpha$ forest covers a large redshift span providing an unique opportunity to constrain process of reionization \citep{2010AAS...21546012P, becker_2015, eilers_2018, bosman1_2018, gaikwad2019, gaikwad2020, bosman_2022, mishra_2022}, 3D distribution of matter in the Universe \citep{petitjean_1995, gatzanaga_1999, Eisenstein_2005, Cole_2005, lymas, lymas2} and constrain nature of dark matter, such as warm dark matter (WDM) \citep{hansen_2002, Viel_2005, garzilli_2021, irsic_17b, palanque_2020}, non-thermal dark matter \citep{baur_2017}, dark matter interactions \citep{bose_2019}, as well as other beyond $\Lambda$CDM models \citep{garzilli_2019, pedersen_2020, sarkar_2021}. In principle, the Ly$\alpha$ forest is a one dimensional probe of cosmological density, temperature and velocity field along a sightline. With the advent of SDSS eBOSS \citep{schelgel_2015, eboss_2016, blomqvist_2019, eboss_2021, nilipour_2022} (and upcoming DESI \citep{ribera_2018, karacayli_2020, walther_2021, desi_2022, satya_2022}, WEAVE \citep{dalton_2012, weave_2022} surveys), it will be possible to observe $\gtrsim 10000$ spectra from fainter, more numerous background quasars in large cosmological volumes $\gtrsim 1 \: h^{-1}$ cGpc.

The interpretation and measurements of cosmological and astrophysical parameters from such a large number of observed spectra would require a careful modelling of the Ly$\alpha$ forest in the cosmological simulations. State-of-the art hydrodynamical simulations, while more accurate and including most of the relevant physics to model Ly$\alpha$ forest, are computationally expensive, limiting their ability to explore parameter space \citep{viel_2006, john_2007, lukic_2015,gaikwad2018, walther_2021}. This is mainly because the dynamic range needed to simulate the Ly$\alpha$ forest that matches observations would require $\sim 1000$ times more grid cells/particles than  current simulations. Even if one develops such simulations, the number of realizations required to capture cosmic variance and to estimate the covariance matrices would be large. Furthermore, adding new physical effects in such simulations while maintaining the code scalability is challenging. Hence one needs efficient semi-numerical methods that may be less accurate but are efficient, flexible and capture the essential physics of the IGM.

Various semi-numerical methods have been developed in the past to efficiently simulate the Ly$\alpha$ forest, with potential applications for parameter space exploration. These methods include (i) simulating dark matter (DM), assuming baryons trace dark matter \citep{petitjean_1995, croft_1998, Sorini_2016}, (ii) simulating few handful of full hydrodynamic simulations for parameters corresponding to a ``best-guess'' model and Taylor expanding the observables around those best-guess values \citep{viel_2006, Pedersen_2021, walther_2021}, (iii) running a large number of inexpensive simulations (e.g., hydro-particle-mesh (HPM)) on a parameter grid and calibrating them using a small number of full hydrodynamic simulations \citep{mcd_2006} (iv) semi-numerical modelling using lognormal simulations \citep{cj91,gh96,bd97,cps01,csp01,viel+02, mattarrese_2002, qiu_2006, Farr_2020}, (v) mapping the baryonic fields from high resolution small box simulation to low resolution large simulation \citep{Borde_2014, lymas, lymas2}. The density and velocity distributions in these models are either generated using cosmological N-body simulations or using physically motivated approximations. 
All these models show that the small scale structure in Ly$\alpha$ forest is sensitive to photo-heating of IGM due to reionization \citep{rorai_2017, bruno_2021}, pressure smoothing effects \citep{peeples_2010, peeples2_2010, rorai_2013, rorai_2017Sci} and suppression in power due to nature of dark matter such as warm dark matter (WDM) \citep{hansen_2002, Viel_2005, garzilli_2021, irsic_17b, palanque_2020}, non-thermal dark matter \citep{baur_2017}, dark matter interactions \citep{bose_2019}, as well as other beyond $\Lambda$CDM models \citep{garzilli_2019, pedersen_2020, sarkar_2021}. 

In order to measure a parameter of interest, one usually needs to marginalize the likelihood over other parameters. For example, constraining the mass of warm dark matter (WDM) \citep{Viel_2005}, fuzzy dark matter \citep{wayne_2000, marsh_2016}, axions \citep{Zhang_2018} or neutrinos \citep{Boyarsky_2019, zelko_2022} requires the marginalization over thermal parameters and pressure smoothing effects in IGM. Furthermore other types of dark matter candidates such as ballistic dark matter (BDM) \citep{Das_2019}, dark photon models \citep{caputo_2021}, self interacting dark matter (SIDM) \citep{carlson_1992, tulin_2018}, late forming dark matter (LFDM) \citep{das_2021}, as well as models within ETHOS (Effective Theory of Structure formation) framework \citep{racine_2016, vogelsberger_2016} are yet to be coupled to hydrodynamic simulations, although see \citet{robles_2017}.

Motivated by this, we develop an end-to-end MCMC analysis method to constrain the astrophysical and cosmological parameters using the lognormal approximation. The lognormal simulations allow us to efficiently generate Ly$\alpha$ absorption spectra on-the-fly in MCMC chains and easily explore a large parameter space. Our approach eliminates the interpolation of Ly$\alpha$ forest statistics that has been widely used in the literature \citep{Bird_2019}. The detailed hydrodynamical simulations of IGM have shown that most Ly$\alpha$ regions are produced by either linear or weakly non-linear regimes. The lognormal model takes into account these non-linearities (to some extent) by assuming number density distribution of the baryons, $n_{\textrm{b}}(x,z)$, to be a lognormal random field. The lognormal model has some interesting features \citep{bd97}, e.g., (i) irrespective of the value of $\delta^L_{\textrm{b}}(x,z)$, $\delta^L_{\textrm{b}}(x,z)$ being the linear baryonic density contrast along line of sight, matter density is always positive, (ii) for low density regions ($\delta^L_{\textrm{b}} \ll 1$), matter density approaches linear theory, $n_{\textrm{b}}(x,z) \propto 1 + \delta^L_{\textrm{b}}(x,z)$.

In this work, we demonstrate the robustness of lognormal model by recovering the thermal and ionization parameters of the IGM from a high resolution smooth particle hydrodynamical simulation with a large dynamic range (SPH hereon). Although we mainly focus on recovering the thermal and ionization parameters of the IGM in this paper, the method can be easily extended to the cosmological parameters. The paper is organized in following way. In \cref{sec:prep}, we discuss the theoretical framework of the lognormal model and our method of generating the Ly$\alpha$ forest spectra. We have also briefly described self-consistent Sherwood simulation suite that are used as a fiducial model. In \cref{sec:method}, we describe our methodology for calculating the flux statistics, covariance matrices and performing likelihood analysis. In \cref{sec:result}, we present our main results of recovering the IGM parameters. We have performed tests studying the effect of box sizes, mass resolution, shape of smoothing filter, Signal-to-Noise ratio (SNR) and initial conditions on recovery of the parameters. Finally, we conclude and discuss applications our method to future work in \cref{sec:conclude}.

\section{Simulations}
\label{sec:prep}
\noindent

In this section, we describe the semi-numerical simulations of the Ly$\alpha$ forest based on the lognormal model, the Sherwood SPH simulations used for comparison and the procedure for parameter recovery using likelihood analysis.
Throughout this work, we fix cosmological parameters for lognormal to Planck 2014 cosmology, the same being used in Sherwood simulations, \{$\Omega_m = 0.308$, $\Omega_{\Lambda} = 1 - \Omega_m$, $\Omega_b = 0.0482$ $h = 0.678$, $\sigma_8 = 0.829$, $n_s = 0.961$, $Y = 0.24$\}, consistent with the constraints from \citet{Planck_2014}. 

\subsection{Semi-numerical simulations based on the lognormal model}

In our framework, the linearly extrapolated power spectrum of DM density field, $P_{\mathrm{DM}}(k)$, is calculated for given set of cosmological parameters.\footnote{In this work, CAMB transfer function to calculate linear matter power spectrum, same as Sherwood simulations \citep{bolton+17-sherwood}.}

The 3D power spectrum of the baryonic density fluctuations at any given redshift $z$ is given by\footnote{Unlike in some literature \citep{Kulkarni_2015, rorai_2017}, where smoothing is done on the Ly$\alpha$ transmitted flux, we use a more physical way by smoothing the DM density field itself.}
\begin{equation}
    P_{\mathrm{b}}(k, z) = D^2(z) P_{\mathrm{DM}}(k)~\mathrm{e}^{-2 x_{\mathrm{J}}^2(z) k^2}.
    \label{eq:Pb_PDM_gaussian}
\end{equation}
where $D(z)$ is the linear growth factor and $x_{\mathrm{J}}(z)$ is the Jeans length. The above relation is based on the assumption that the baryonic fluctuations follow the dark matter at large scales $k^{-1} \gg x_{\mathrm{J}}$ and are smoothed because of pressure forces at scales $k^{-1} \lesssim x_\mathrm{J}$. The default form of the smoothing function is taken to be Gaussian, as shown by \citet{gnedin+03, lukic_2015} that a Gaussian smoothing filter does a better job in reproducing the linear baryonic fluctuations at transitional regions between large and small scales. However, we explore a variant where the function is Lorentzian \citep{1993ApJ...413..477F}. (see \cref{sec:result}).
 
Since the Ly$\alpha$ forest probes the cosmic fields only along the lines of sight, it is sufficient to generate the baryonic density field $\delta_{\mathrm{b}}^L(x, z)$ and the corresponding line of sight component of the velocity fields $v_{\mathrm{b}}^L(x, z)$ only along one direction.

We generate the realizations of these two fields accounting for the cross correlation between them; we refer the readers to \citet{cj91,gh96,bd97,cps01,csp01,viel+02,mattarrese_2002} for full details of the method. The fields $\delta_{\mathrm{b}}^L(x, z)$ and $v_{\mathrm{b}}^L(x, z)$ thus generated are based on the linearly extrapolated power spectra and are Gaussian random fields. 
 
 To account for the quasi-linear description of the density field, we employ the lognormal assumption and take the baryonic number density to be
 \begin{equation}
     n_{\mathrm{b}}(x,z) = A~\mathrm{e}^{\delta^L_{\mathrm{b}}(x,z)},
 \end{equation}
 where $A$ is a normalization constant fixed by setting the average value of $n_{\mathrm{b}}(x,z)$ to the mean baryonic density $\bar{n}_\mathrm{b}(z)$ at that redshift.

At this point, we should stress that the simple lognormal assumption for the baryonic field is \emph{not} a good description of that found in the SPH simulations (we have independently verified this). Our aim is not to develop a model to describe all properties  of baryonic matter in the IGM, rather we want to find a quick and approximate way of generating the Ly$\alpha$ forest that can describe the SPH data for a similar set of IGM parameters. As is known, and also we shall discuss later, the Ly$\alpha$ transmitted flux is a highly non-linear (and to some extent non-local) function of the baryonic density, hence it might be possible to find a simple model which agrees with the Ly$\alpha$ flux statistics of the SPH simulations, even though the description of the baryonic field in general might be inaccurate. Whether the model works in this respect would form the basis of the results presented.

We calculate the neutral hydrogen number density assuming photo-ionization equilibrium as,
\begin{equation}
     \alpha_A[T(x,z)]~n_{\mathrm{p}}(x,z)~n_{\mathrm{e}}(x,z) = n_{\mathrm{HI}}(x,z)~\Gamma_{\mathrm{HI}}(z),
 \end{equation}
 where $\alpha_A(T)$ is the recombination coefficient at temperature $T$ (taken to be of A-type in this work, appropriate for the low-density IGM), $n_\mathrm{p}, n_\mathrm{e}$ are the number densities of protons and free electrons respectively and $\Gamma_{\mathrm{HI}}$ is the hydrogen photoionization rate (assumed to be homogeneous). Assuming a fully ionized IGM, $n_\mathrm{p}, n_\mathrm{e}$ are given by,

 \begin{equation}
     n_p(x,z) = \frac{4(1 - Y)}{4 - 3Y}n_{\textrm{b}}(x,z)\, ; \, n_e = \frac{4 - 2Y}{4 - 3Y}n_{\textrm{b}}(x,z)
 \end{equation}where $Y (\sim 0.24)$ is helium weight fraction.

The temperature field is computed assuming a power-law relation with the baryon overdensity $\Delta_\mathrm{b}$
 \begin{equation}
    T(x, z) = T_0(z) \left[\Delta_\mathrm{b}(x,z)\right]^{\gamma(z)-1}
    = T_0(z) \left(\frac{n_\mathrm{b}(x,z)}{\bar{n}_b(z)}\right)^{\gamma(z)-1},
    \label{eq:T_Delta}
\end{equation}
where $T_0$ is the temperature at mean baryonic density and $\gamma$ is the slope of the power-law relation \citep{gh96}.

With the above relations in hand, we generate one dimensional fields at a chosen central redshift $z$, for a given length of the skewer, $z\, \epsilon \, [2.475, 2.525]$ and a pixel size equivalent to $v_{\textrm{res}} = 1.95$ km s$^{-1}$. The fields are generated on a uniform grid in the comoving position $x$ with the number of grid points determined by the box size and the resolution. The Ly$\alpha$ optical depth at each of these grid points $x_i$ can then be calculated as
\begin{align}
    \tau(x_i, z) &= \frac{c I_{\alpha}}{\sqrt{\pi}} \sum_j \delta x \frac{n_{\mathrm{HI}}(x_j,z)}{b(x_j,z)[1+z(x_j)]}
    \notag \\
    &\times  V_{\alpha}\left(\frac{c[z(x_j)-z(x_i)]}{b(x_j,z)[1+z(x_i)]}+\frac{v^L_{\mathrm{b}}(x_j,z)}{b(x_j,z)}\right),
\end{align}
where $\delta x$ is the separation between the grid points (i.e., the grid size), $I_{\alpha} = 4.45\, \times 10^{-18}$ cm$^2$ is the Ly$\alpha$ absorption cross section and $V_{\alpha}(\Delta v / b)$ is the Voigt profile for the Ly$\alpha$ transition. The summation is over all grid points along the skewer. The thermal velocity, which measures the thermal width of the absorption lines, is given by
\begin{equation}
    b(x,z) = \sqrt{\frac{2 k_{\mathrm{boltz}} T(x,z)}{m_{\mathrm{p}}}},
\end{equation}
where $m_\mathrm{p}$ is the proton mass. The comoving positions $x_i$ can be converted to a redshift $z(x_i)$ using the usual (implicit) relation
\begin{equation}
    x(z) = \int_0^z \frac{c~\mathrm{d}z'}{H(z')},
\end{equation}
where $H(z)$ is the Hubble parameter. We have also applied periodic boundary conditions while computing optical depths.

The observable in Ly$\alpha$ forest spectra is the transmitted flux defined as
\begin{equation}
    F(x_i, z) = \mathrm{e}^{-\tau(x_i, z)}.
\end{equation}

For calculating the observables, the independent variable is transformed from the comoving position $x_i$ to the velocity $v_i$ \citep{Kim_2004}.
\begin{equation}
    v_i = H(z)~x_i.
\end{equation}

In our lognormal framework of the Ly$\alpha$ forest, there are 4 main free parameters: 
\begin{itemize}
    
    \item $x_{\mathrm{J}}$: the Jeans length (see equation~\ref{eq:Pb_PDM}) which controls the pressure smoothing of the baryonic density field.  
    
    \item $\Gamma_{12}$: the photoionization rate $\Gamma_\mathrm{HI}$ in units of $10^{-12}$s$^{-1}$. 
    
    \item $T_0$ and $\gamma$: parameters characterizing the temperature-density relation, see equation (\ref{eq:T_Delta}). 

\end{itemize}

Note that the above parameters can be redshift-dependent; since we are concerned only with a single redshift in this work, they can be treated as constants. 

\subsection{Sherwood simulations}
\noindent

We use publicly available Sherwood simulations \citep{bolton+17-sherwood} suite that were performed with a modified version of the cosmological smoothed particle hydrodynamics code P-Gadget-3, an extended version of publicly available GADGET-2 code \citep{Springel_2005}\footnote{\url{https://wwwmpa.mpa-garching.mpg.de/gadget/}}. The Sherwood suite consists of cosmological simulation boxes with volume
ranging from $10^3$ to $160^3$ $h^{-3}\,\textrm{cMpc}^3$ and contains number particles ranging from $2\times 512^3$ to $2 \times 2048^3$. The size and resolution of simulation box are suitable for
studying the small scale structures probed by Ly$\alpha$ forest.
The properties of Ly$\alpha$ forest from Sherwood simulation suite are well converged \citep{bolton+17-sherwood}. 

As the default, we choose a box of volume $40^3$ $h^{-3}\,\textrm{cMpc}^3$ containing $2 \times 2048^3$ particles (we will refer to the box as 40-2048). The box is chosen such that the the large scale modes are captured while at the same time has a resolution appropriate for Ly$\alpha$ forest studies \citep{bryan_1999, Meiksin:2000eg, mcd_2003}. We choose the default redshift as $z = 2.5$. The redshift has been chosen so as to avoid the complications arising from shock heating (helium reionization) at lower (higher) redshifts. We will study the comparison of lognormal model with SPH simulations at other redshifts in a different work. The simulation box size at this redshift corresponds to a redshift path length of $\Delta z \sim 0.0497$. In addition to the default box, we have used other boxes (40-1024 and 80-2048) for studying the convergence of our results with respect to resolution and box size, results of which we report in later sections. For each of these models, 5000 random skewers are extracted and the Ly$\alpha$ optical depths are calculated .

We use Sherwood simulation suite as true or fiducial model to recover the four free parameters using our lognormal model.
The relevance of the four free parameters of the lognormal model (as described in the previous section) in the context of these detailed simulations is as follows,

\begin{itemize}
    \item $x_{\mathrm{J}}$: the Jeans length, which characterizes the pressure smoothing, cannot be represented by a single parameter in the simulations. The pressure smoothing is different in different locations in the simulation volume and is determined by the past thermal history of the location under consideration \citep{rorai2_2017,gaikwad2018, walther_2019, nasir_2020}. Representing this complex physics with a single free parameter is an obvious simplification in the lognormal model.
    
    \item $\Gamma_{12}$: the photoionization rate is taken from \citep{haardt_2012} assumed to be homogeneous and is externally supplied to the SPH simulations. However the $\Gamma_{12}$ in our lognormal model is a free parameter. For the default simulation box, the value turns out to be $\Gamma_{12} = 0.984$.
    
    \item $T_0$ and $\gamma$: the two thermal parameters are not input parameters in the SPH simulations as the temperature is determined by solving the appropriate temperature and ionization evolution equations. However, these parameters can be derived from the $T - \Delta_{\mathrm{b}}$ phase space diagram. We have checked and found that the $T - \Delta_{\mathrm{b}}$ relation in the low-density IGM is indeed well represented by a power-law. We fit a straight line to the $\log T - \log \Delta_{\mathrm{b}}$ points and infer $T_0 = 1.19 \times 10^4$~K and $\gamma = 1.56$ for the default simulation box.
\end{itemize}

In the next section, we described our method of recovering the four parameters using MCMC analysis.

\section{Method}
\label{sec:method}
\subsection{Ly$\alpha$ Forest Statistics}
\noindent

In this section, we describe the method to compare the outputs from SPH and lognormal models using likelihood analysis. Observations of the flux spectra are affected by line spread function (LSF) of the instrument  as well as noise. In order to incorporate these noise properties we first add these observational effects to the both SPH and lognormal flux spectra. We first convolve the transmitted flux field with a Gaussian LSF having Full Width at Half Maximum (FWHM) of 7 km s$^{-1}$, similar to spectral resolution of current instruments \citep{bolton+17-sherwood}. We then add a Gaussian random noise of  SNR per pixel = 50 \citep{bolton+17-sherwood}, typical of high quality data \citep{L_pez_2016, Murphy_2018, O_Meara_2020}. Our simplistic treatment of noise addition leads to almost noise-less spectra at pixels where $F \sim 0$. To avoid this, we also add a noise floor ($=0.01$), which gives the minimum standard deviation of the Gaussian noise per pixel. Note that we treat spectra from both the SPH and lognormal models identically in this regard.

In this work, we consider several statistics to compare the lognormal model with the SPH simulations. These are:

\begin{enumerate}
    \item The flux probability distribution (FPDF): in this case, we bin all the flux values $F$ and compute the probability distribution $\mathrm{d}P / \mathrm{d}F$. We calculate FPDF from $F = 0$ to $F = 1$ in 11 bins with a bin width, $dF = 0.1$ (which is sufficiently wide for SNR = 50 so as to avoid any effects of the noise). Because of expected sky subtraction uncertainty (continuum placement uncertainty) at low (high) flux values, we put all pixels with $F \leq 0.05$ ($F \geq 0.95$) in the first (last) bin \citep{gaikwad2017a,gaikwad2017b}. Note that the SPH does not actually have sky subtraction and it is done merely to mimic observed data. We use all the available pixels to normalize the FPDF, but use bins only in the range $0.1 \leq F \leq 0.8$ for the likelihood analysis because the flux at $F < 0.1 (> 0.8)$ could be dominated by sky subtraction (continuum placement uncertainty) \citep{prakash21}. The one-point statistics was first studied by \citet{jenkins_1991} and is sensitive to thermal state of IGM ($T_0$ and $\gamma$) \citep{becker_2007, bolton_2008, 10.1111/j.1365-2966.2009.14914.x, McQuinn_2009, calura_2012, Lee_2015}.
    \item The mean flux ($\bar{F}$): this is simply the flux averaged over all pixels under consideration. Strictly speaking, this quantity is not independent of the full FPDF, however, we can treat it as an independent observable since we consider only a subset of FPDF bins in our analysis. $\bar{F}$ is also sensitive to photo-ionization coefficient $\Gamma_{12}$ \citep{Tytler_2004, Bolton_2005, viel_2009}. Below, we account for the correlation between FPDF and $\bar F$ when computing the covariance matrix.
    \item The flux power spectrum (FPS): we use the flux contrast field $\delta_F \equiv F / \bar{F} - 1$, take its Fourier transform and calculate the power spectrum \citep{Zaldarriaga_2001, Nasir_2016, mishra_2022}. We consider modes from $k \geq 0.005$ s km$^{-1}$ to $k \leq 0.164$ s km$^{-1}$ in 11 equally spaced logarithmic bins. The minimum value of $k$ chosen is $\sim 3$ times $\frac{2 \pi}{L}$ so as to avoid any effects arising from the finite box size. The maximum value of $k$ is $\sim 0.2$ times the Nyquist frequency to avoid aliasing. We also make sure that the maximum value of $k$ does not venture into regions where the power spectrum becomes noise dominated. FPS is sensitive to both thermal state and Jeans length at small scales \citep{viel2_2006}.
\end{enumerate}

\subsection{Covariance Matrix Calculations}
\noindent

For the statistical analysis, we need to compute not only the above statistics but also the errors on them. While computing these errors in actual observational data can be quite non-trivial because of a limited number of sightlines, the situation is somewhat straightforward in the case of simulations where one has access to several realizations of the sample. Let us first discuss the method to compute the errors in the case of SPH simulations. We first define the ``sample'' as the set of spectra for which the total path length $DX = 6.2$, similar to the available observational data at $z \sim 2.5$ \citep{bolton+17-sherwood}. The path length for one sightline can be related to the redshift interval $\Delta z$ as
\begin{equation}
    \Delta X = \frac{H_0(1 + z)^2}{H(z)}\Delta z
\end{equation}
For the simulation box length of $40 h^{-1}$~cMpc, this length corresponds to $N_\mathrm{spec} = 40$ sight lines. We take randomly chosen $N_\mathrm{spec}$ sight lines from the box which corresponds to the ``sample''. The statistics computed for this sample would be the ``data points'' used for comparing with the lognormal model.

To compute the errors on SPH statistics, we consider $N$ realizations of this sample each consisting of $N_\mathrm{spec}$ sight lines. Since we have access to 5000 sight lines in the box, this fixes $N = 125$ for this work. Since 125 realizations are not enough to properly calculate covariance matrix, we have used Jackknife resampling for the said purpose. To obtain Jackknife samples, we take a chunk of $N_\mathrm{spec}$ (=40 in this case) sightlines and average over these sightlines by leaving one sightline out. This gives us $N_\mathrm{spec}$ Jackknife samples, each averaged over $N_\mathrm{spec} - 1$ sightlines. We use these 40 samples to calculate covariance matrix. We repeat this exercise in each of the $N$ (=125 in this case) chunks to obtain $N$ covariance matrices. The SPH covariance matrix is the mean of these $N$ covariance matrices. Furthermore, it is possible that there exists non-trivial correlations between these statistics and hence we will take into account the full covariance matrix which involves all the data points. To do this, let us define the full data vector as $X_i = \left\{\mathrm{d}  P(F_i) / \mathrm{d} F, \bar{F}, P_F(k_i)\right\}$. The covariance matrix for the sight lines in the SPH simulation can be written as
\begin{multline}
    C^{\mathrm{SPH}, (n)}(i,j) = \frac{N_\mathrm{spec}-1}{N_\mathrm{spec}} \sum_{k=1}^{N_\mathrm{spec}} \left[\tilde{X}^{(n)}_{-k,i} - \bar{\tilde{X}}^{(n)}_{i}\right] \\ \left[\tilde{X}^{(n)}_{-k,j} - \bar{\tilde{X}}^{(n)}_{j}\right],
    \label{eq:cov_sph_jkn}
\end{multline}
where $C^{\mathrm{SPH}, (n)}(i,j)$ is the Jackknife covariance matrix in $n^{th}$ chunk. Also
\begin{equation*}
    \tilde{X}^{(n)}_{-k,i} = \frac{1}{N_\mathrm{spec}-1} \sum_{m=1,\neq k}^{N_\mathrm{spec}} X_i^{(n,m)}
\end{equation*}
\begin{equation*}
    \bar{\tilde{X}}^{(n)}_{i} = \frac{1}{N_\mathrm{spec}} \sum_{k=1}^{N_\mathrm{spec}}\tilde{X}^{(n)}_{-k,i}
\end{equation*}
and $X^{(n,m)}_i$ is the statistic in the $i^{th}$ bin of $m^{th}$ sightline in $n^{th}$ chunk of the sample. Therefore, the SPH covariance matrix is given by,
\begin{equation}
    C^{\mathrm{SPH}}(i,j) = \frac{1}{N}\sum_{n=1}^N C^{\mathrm{SPH}, (n)}(i,j)
     \label{eq:cov_sph}
\end{equation}
The indices $i$ and $j$ take values from 1 to the number of bins used in the analysis (8 + 1 + 11 in this work).

Since the lognormal model too is stochastic, we need to account for this in the variance of the data points. In order to keep the computing time under control, we generate 40000 sight lines for the lognormal and then divide them into $N = 1000$ realizations of samples each containing $N_\mathrm{spec} = 40$ sight lines. The calculation of the covariance matrix $C^\mathrm{LN}(i,j)$ is given by
\begin{equation}
    C^\mathrm{LN}(i,j) = \frac{1}{N-1} \sum_{n=1}^{N} \left[\tilde{X}^{(n)}_{i} - \bar{\tilde{X}}_{i}\right] \left[\tilde{X}^{(n)}_{j} - \bar{\tilde{X}}_{j}\right],
    \label{eq:cov_ln}
\end{equation}
To calculate covariance matrix for lognormal, we use Planck 2014 cosmology and fix \{$T_0$, $\gamma$, $\Gamma_{12}$\} to true / inferred values used in SPH. Unlike SPH, which uses pixel dependent pressure smoothing scheme, lognormal employs a single parameter $x_J$. Therefore, $x_J$ has no "true" value. Therefore, to calculate lognormal covariance matrix, we use $x_J = 0.12$ \Mpch, a rather arbitrary value. We have verified that even though the covariance matrices are parameter dependent, MCMC chains are not. The total covariance is simply the sum of the two, i.e.,
\begin{align}
    C^\mathrm{tot}(i,j) &= C^\mathrm{SPH}(i,j) +  C^\mathrm{LN}(i,j)
    \label{eq:cov_tot}
\end{align}

\begin{figure*}
\centering
\includegraphics[width=1\textwidth]{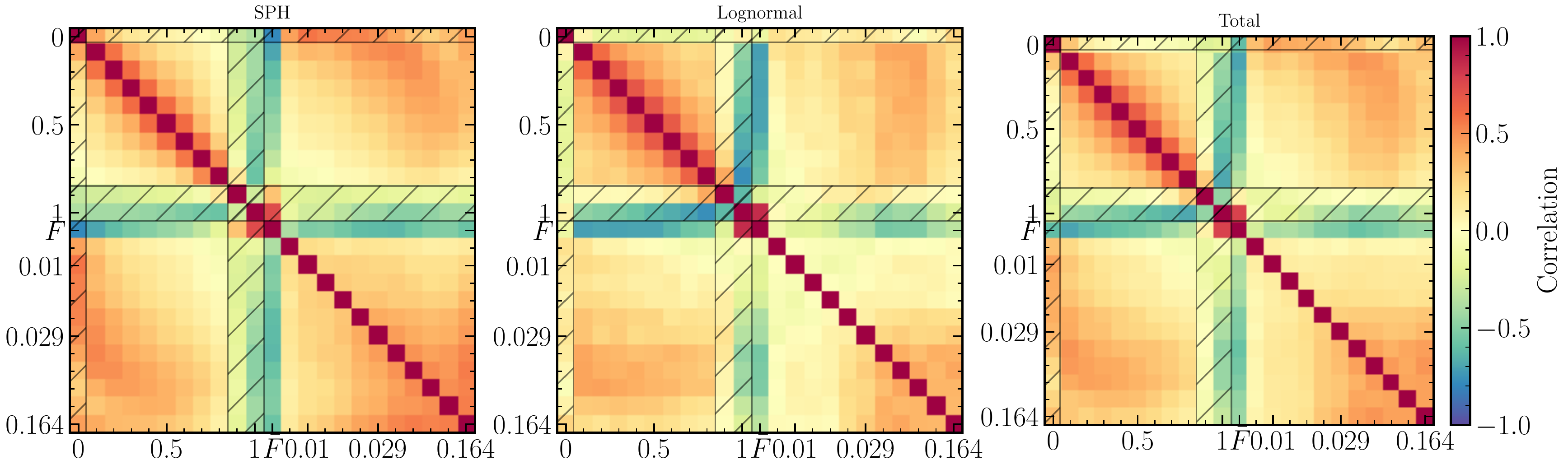}
\caption{Correlation matrices with cross-correlations between FPDF, $\bar{F}$ and FPS for default case for SPH, lognormal and total. Shaded regions are pixels that are not included in likelihood analysis. Lognormal shows significantly higher (lower) positive correlations in FPDF (FPS) compared to SPH. Both SPH and lognormal show negative correlations between $\bar{F}$ and FPS, although the correlations in lognormal occur at small scales while in SPH it is prevalent at all scales.}
\label{fig:cov}
\end{figure*}

For each of the covariance matrices $C(i,j)$, we can calculate the corresponding correlation matrices as
\begin{equation}
    \mathrm{Corr}(i,j) = \frac{C(i,j)}{\sqrt{C(i,i)~C(j,j)}}.
\end{equation}
Fig. \ref{fig:cov} shows the SPH, lognormal, and total correlation matrices. Corresponding to the FPDF, the neighbouring bins are positively correlated and this correlation fades away as we consider far away bins. We also see that bins in the lognormal model show stronger correlations compared to those for the SPH simulations.

For the FPS, due to non-linearities at small scales, modes show strong positive correlations while the large scale modes are relatively weakly correlated. Since the treatment of the non-linearities in the SPH simulations is more accurate than in the lognormal model, we see that correlations extend up to larger scales in SPH compared to lognormal.

\subsection{Likelihood Analysis}
\noindent

For the likelihood analysis, we use the publicly available Markov Chain Monte Carlo (MCMC) code \texttt{cobaya} 
\citep{Lewis_2002}\footnote{\url{https://cobaya.readthedocs.io/en/latest/sampler_mcmc.html}} and compute the posterior distribution of the free parameters of the lognormal model by comparing with the SPH simulations. The main ingredient required for the analysis is the $\chi^2$ which is calculated using
\begin{align}
    \chi^2 &= \left(\mathbf{X}^{\mathrm{LN}} - \mathbf{X}^{\mathrm{SPH}}\right) \left(\mathbf{C}^{\mathrm{tot}} \right)^{-1} \left(\mathbf{X}^{\mathrm{LN}} - \mathbf{X}^{\mathrm{SPH}}\right)^T,
\end{align}
where $\mathbf{X}$ are the row vectors corresponding to the data points $X_i$ and $\mathbf{C}$ are the appropriate covariance matrices. We reiterate that, while using $\bar{F}$ as a data point may seem redundant given we are already using FPDF, unlike FPDF which is truncated between $0.1 \leq F \leq 0.8$ for likelihood analysis, $\bar{F}$ is obtained by averaging over all pixels, thereby including information from edges as well. Note that the above form of the $\chi^2$, used in the our default analysis, also accounts for correlation between all statistics. We investigate the effect of ignoring correlation between FPS and \{FPDF+$\bar{F}$\} as a variant in \cref{sec:result}.

We use 32 walkers using 32 processors for each MCMC run. To determine when a chain is converged, we use Gelman-Rubin statistics parameter, $R_{-1} = 0.01$ \citep{gelman_rubin}. The convergence takes $\sim 2$ days on a 2048 grid with a path length of 6.2 at $z = 2.5$. 
All MCMC calculations were performed on the Pegasus cluster at IUCAA.

\section{Results}
\label{sec:result}
\noindent

In this section, we present the recovery of the free parameters of the lognormal by comparing with the SPH simulations.

\subsection{Parameter recovery}
\noindent

\begin{figure}
\centering
\includegraphics[width=0.4\textwidth]{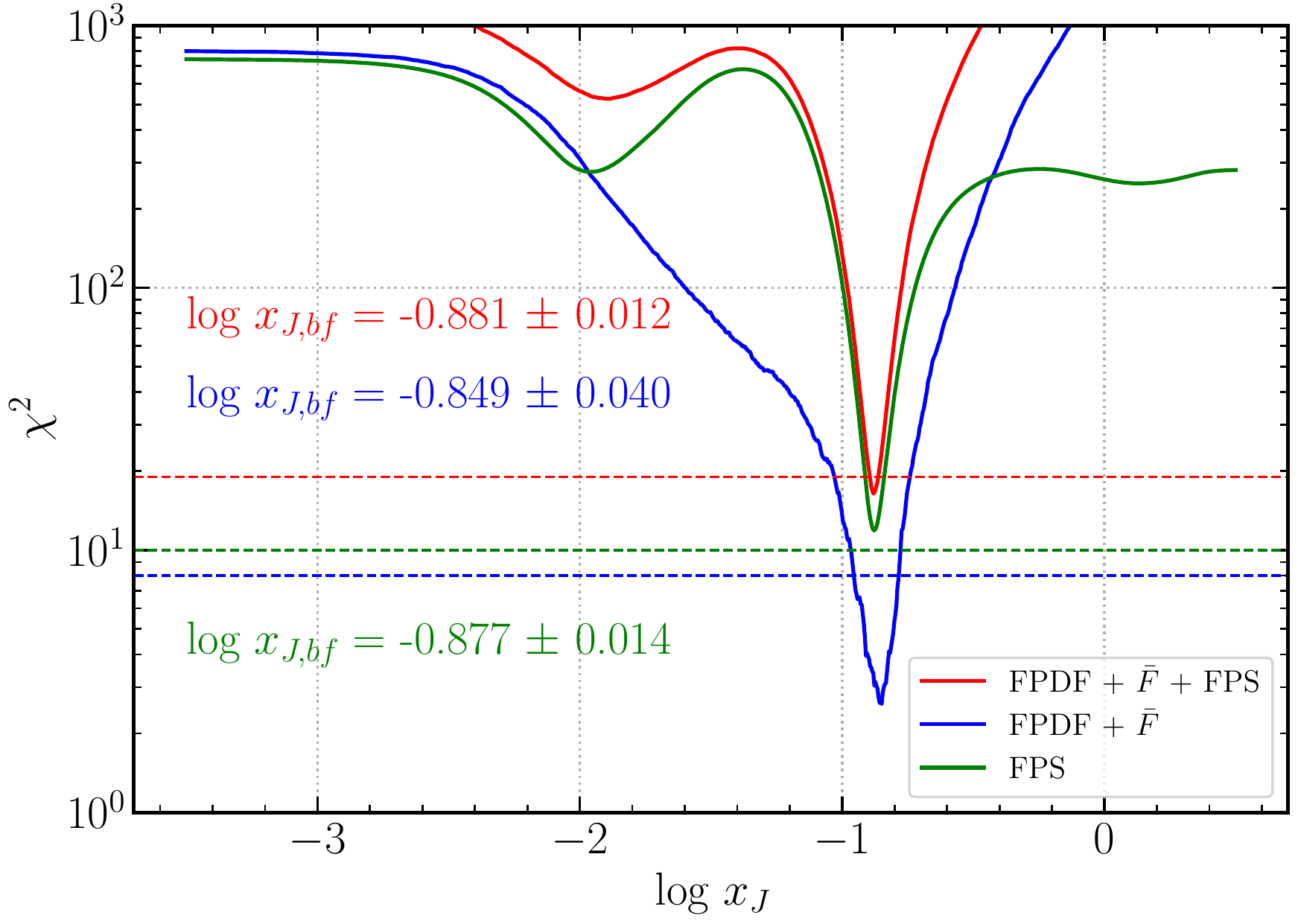}
\caption{The solid line shows $\chi^2$ vs log $x_{\textrm{J}}$ obtained from a simple 1D likelihood analysis (see text for the details). The dashed horizontal lines show the degrees of freedom in each case.}
\label{fig:1d_chi2}
\end{figure}

\begin{figure*}
\centering
\includegraphics[width=1\textwidth]{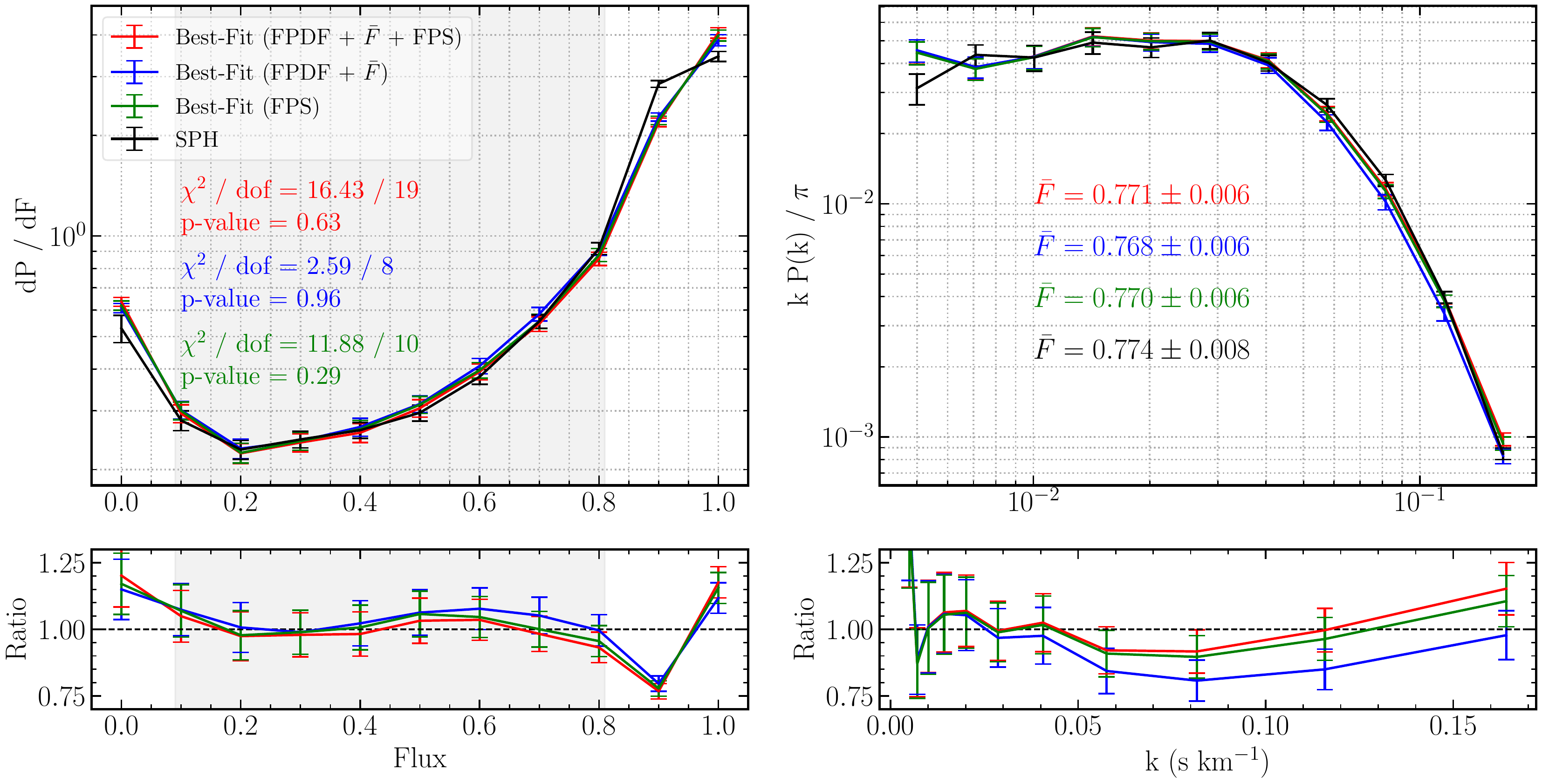}
\caption{We show the best-fit flux statistics obtained from the 1D likelihood analysis (red) and those obtained from the SPH simulation used (black). Top left (right) panel shows FPDF (FPS). Errorbars on FPDF, $\bar{F}$ and FPS are obtained from diagonal elements of covariance matrices. To show the goodness of the fit, we mention the value of the minimum $\chi^2$ per degree of freedom and the corresponding $p$-value in the same panel. In the top right panel, we mention the values of mean flux along with the standard deviation for the best-fit model and SPH in the corresponding colours. Bottom panels show the ratios of flux statistics ($S_{\textrm{LN}}/ S_{\textrm{SPH}}$), where $S$ represents the FPDF (FPS) in left (right) panel. Grey regions in left panels show flux bins used in likelihood analysis.}
\label{fig:1d_chi2_stat}
\end{figure*}

We will first present results for our default configuration, as described in the previous sections. But before embarking on a full MCMC analysis, let us first try to understand what would be the typical value of the Jeans length $x_{\mathrm{J}}$, the parameter which does not have an obvious counterpart in the SPH simulations. Keeping this in mind, we do a simple $\chi^2$-minimization using a 1D grid in $\log$ $x_{\mathrm{J}}$ and find the value of $x_{\mathrm{J}}$ which best fits the simulation output. For the other three parameters, namely, $\Gamma_{12}, T_0, \gamma$, we use the values as in the SPH simulation. As mentioned earlier, we use the three flux statistics FPDF, mean flux and FPS for calculating the $\chi^2$. Additionally, we repeat the exercise by using statistics FPDF + $\bar{F}$ and FPS independently.

The plot of $\chi^2$ as a function of $x_{\mathrm{J}}$ for all three cases is shown in Fig. \ref{fig:1d_chi2}. Using all the three statistics (red curve), we see a clear minimum which provides a best-fit value of $x_{\textrm{J}}$ as 0.13 \Mpch. This value is of the same order as the one obtained by assuming Ly$\alpha$ absorbers to be in hydrostatic equilibrium at a temperature $\sim 10^4$K \citep{schaye_2001}. The minimum $\chi^2$ per degree of freedom $\chi^2_{\textrm{min}} / \mathrm{dof} \lesssim 1$, which implies that the fit is quite good. This can be clearly seen in Fig. \ref{fig:1d_chi2_stat} where we have used the best-fit value of $x_{\textrm{J}}$ to compare the flux statistics of the lognormal model to that of the SPH. All three statistics, FPDF, $\bar{F}$ and FPS for the lognormal model and SPH seem to be in very good agreement.

To understand the effects of each statistic, we also present minimum reduced $\chi^2$, $\chi^2_{\nu, \textrm{min}}$ and best-fit flux statistics obtained from using FPDF+$\bar{F}$ (blue) and FPS (green) independently. We see that in both cases, best-fit value of $x_{\textrm{J}}$ is roughly similar.

Although the simple one-parameter lognormal model produced a good enough fit to the SPH simulations, we next explore the case where all the free parameters of the lognormal are kept free.

\begin{figure*}
\centering
\includegraphics[width=0.8\textwidth]{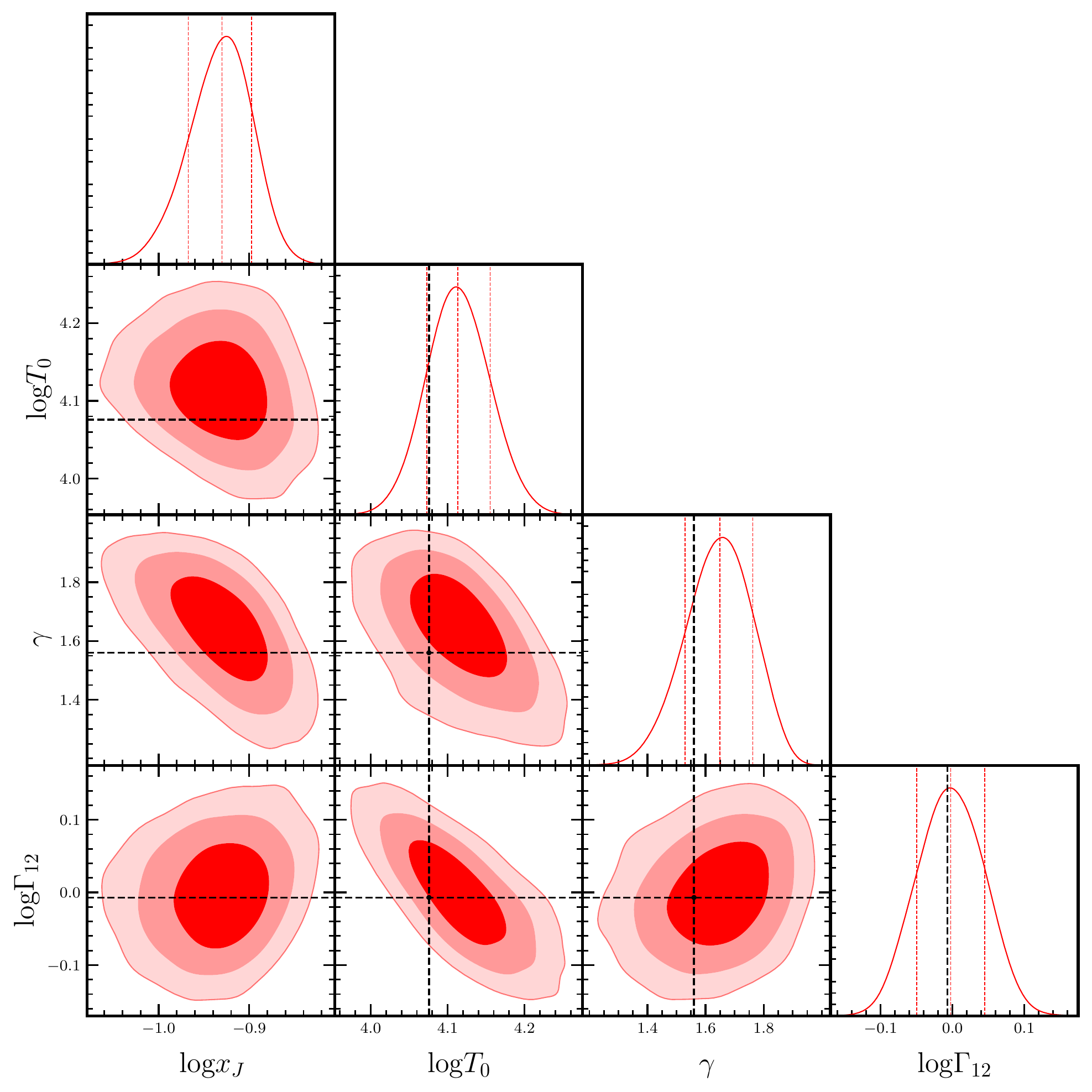}
\caption{Parameter estimates obtained from MCMC runs in a 4D space. Black lines show true / inferred values of IGM parameters in simulated data. We find that the true values of all three IGM parameters, $T_0$, $\gamma$ and $\Gamma_{12}$ are recovered well within $1-\sigma$ from their median. $x_{\textrm{J}}$ is non-degenerate with the other three parameters but the three parameters show some degeneracies among each other.}
\label{fig:corner_main}
\end{figure*}

\begin{figure*}
\centering
\includegraphics[width=1\textwidth]{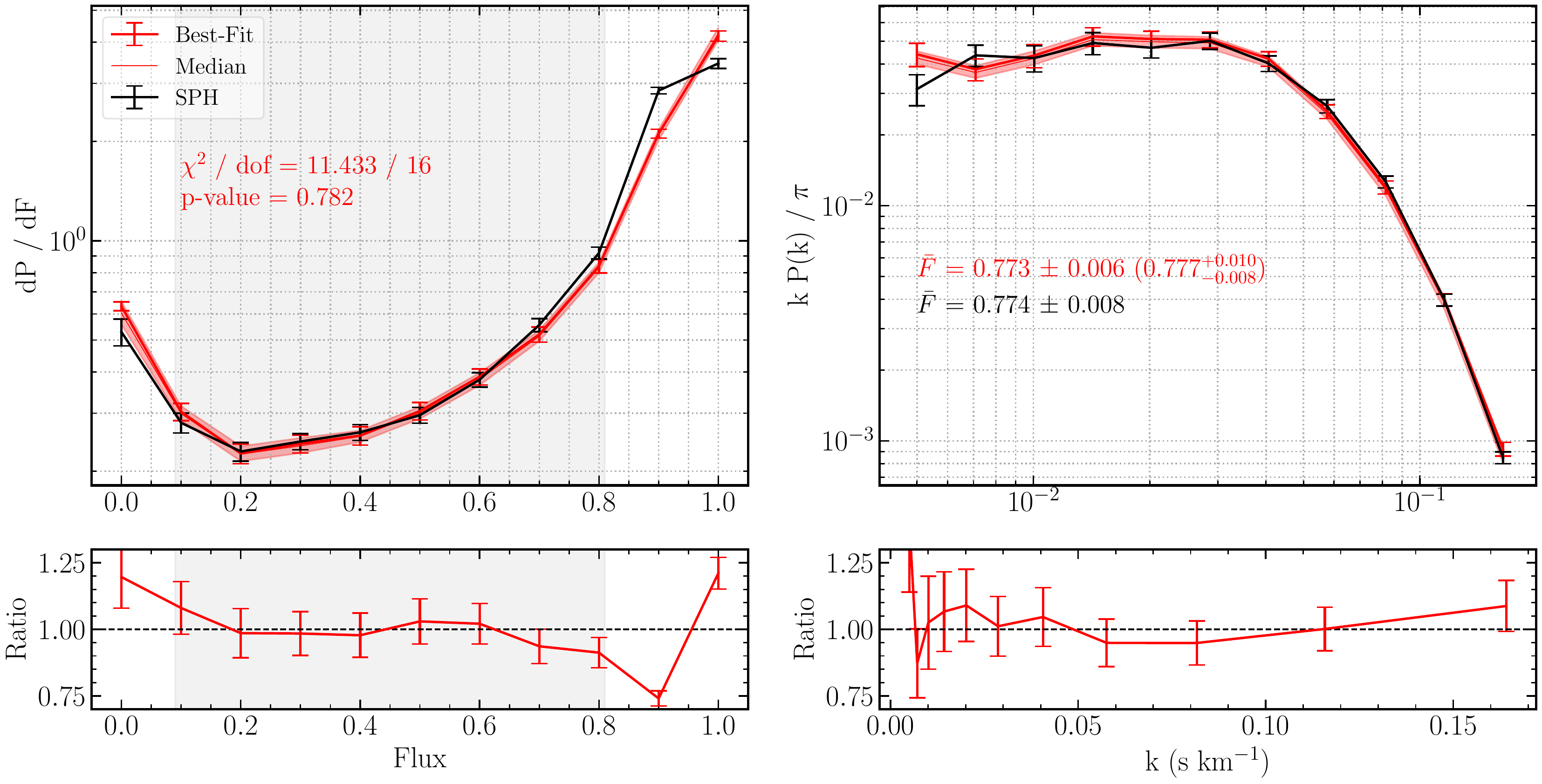}
\caption{Flux statistics for best-fit values of parameters obtained from MCMC run. Fits for all three statistics are in very good agreement with SPH. Banded regions in FPDF and FPS show 16 and 84 percentiles of 1000 randomly selected parameter vectors from MCMC chain. We also mention the similar $1-\sigma$ estimate on mean flux in top right panel (round brackets).}
\label{fig:stat_main}
\end{figure*}

The priors on the parameters, their true values (i.e., the values used in or obtained from the SPH simulations) and the constraints on them from the MCMC runs are reported in Table \ref{table:1}. In figs. \ref{fig:corner_main} and \ref{fig:stat_main} we show the contour plots (68.3, 95.4, 99.7 percentiles) obtained from MCMC run and best-fit flux statistics respectively. From Fig. \ref{fig:corner_main}, it is evident that the lognormal model does a decent job at recovering all three parameters, $T_0$ $\gamma$ and $\Gamma_{12}$.  True values of the parameters lie within $1-\sigma$ from median values in the MCMC chain. We also see $x_{\textrm{J}}$ is non-degenerate with all other three parameters, \{$T_0$, $\gamma$, $\Gamma_{12}$\} while these three parameters themselves are mildly degenerate with each other. These degeneracies may significantly affect parameter estimates. E.g., overestimating value of $T_0$ smooths (or lower) the FPS at small scales. It also reduces regions with flux absorption, thereby lowering FPDF for $0.1 \leq F \leq 0.8$ (range used in likelihood analysis). To counteract this effect, value of $\Gamma_{12}$ is underestimated. This is clearly evident from the strong anti-correlation seen in the log $T_0$ - log $\Gamma_{12}$ histogram in fig. \ref{fig:corner_main}. We can draw similar conclusions for degeneracies among other IGM parameters. Fig. \ref{fig:stat_main} shows best-fit  statistics for 4D parameter estimates (similar to 1D estimates). Again, we see best-fits for FPS agreeing very well with the data but FPDF is consistently overestimated (due to positive correlations between neighbouring bins in FPDF correlation matrix). 

\begin{center}
\begin{table}
\begin{tabular}{||c c c c||} 
 \hline
 Parameter & Prior & True Value & Best-Fit Value \\ [0.5ex]
 \hline\hline
 log $x_{\textrm{J}}$ & [-3.5, 0.5] & - & $-0.921(-0.930)^{+0.033}_{-0.037}$ \\ 
 \hline
 log $T_0$ & [2.5, 5.5] & 4.076 & $4.113(4.113)^{+0.042}_{-0.040}$ \\
 \hline
 $\gamma$ & [-0.5, 5] & 1.56 & $1.616(1.650)^{+0.113}_{-0.121}$ \\
 \hline
 log $\Gamma_{12}$ & [-2, 2] & -0.007 & $-0.002(-0.002)^{+0.047}_{-0.047}$ \\
 \hline
\end{tabular}
\caption{Priors, true and best-fit values for parameter estimates in MCMC run. Values in brackets show median estimates alongwith $1-\sigma$. All priors in this work are uniform.}
\label{table:1}
\end{table}
\end{center}

\subsection{Effect of modeling and observational uncertainties on parameter recovery}
\noindent

We will now discuss effects of varying configurations such as ignoring correlations between FPS and \{FPDF+$\bar{F}$\}, baryon smoothing filter, noise properties, data seed and box size and resolution on the results obtained in the previous section. 

\subsubsection{Separate Covariance}
\noindent

\begin{figure*}
\centering
\includegraphics[width=0.8\textwidth]{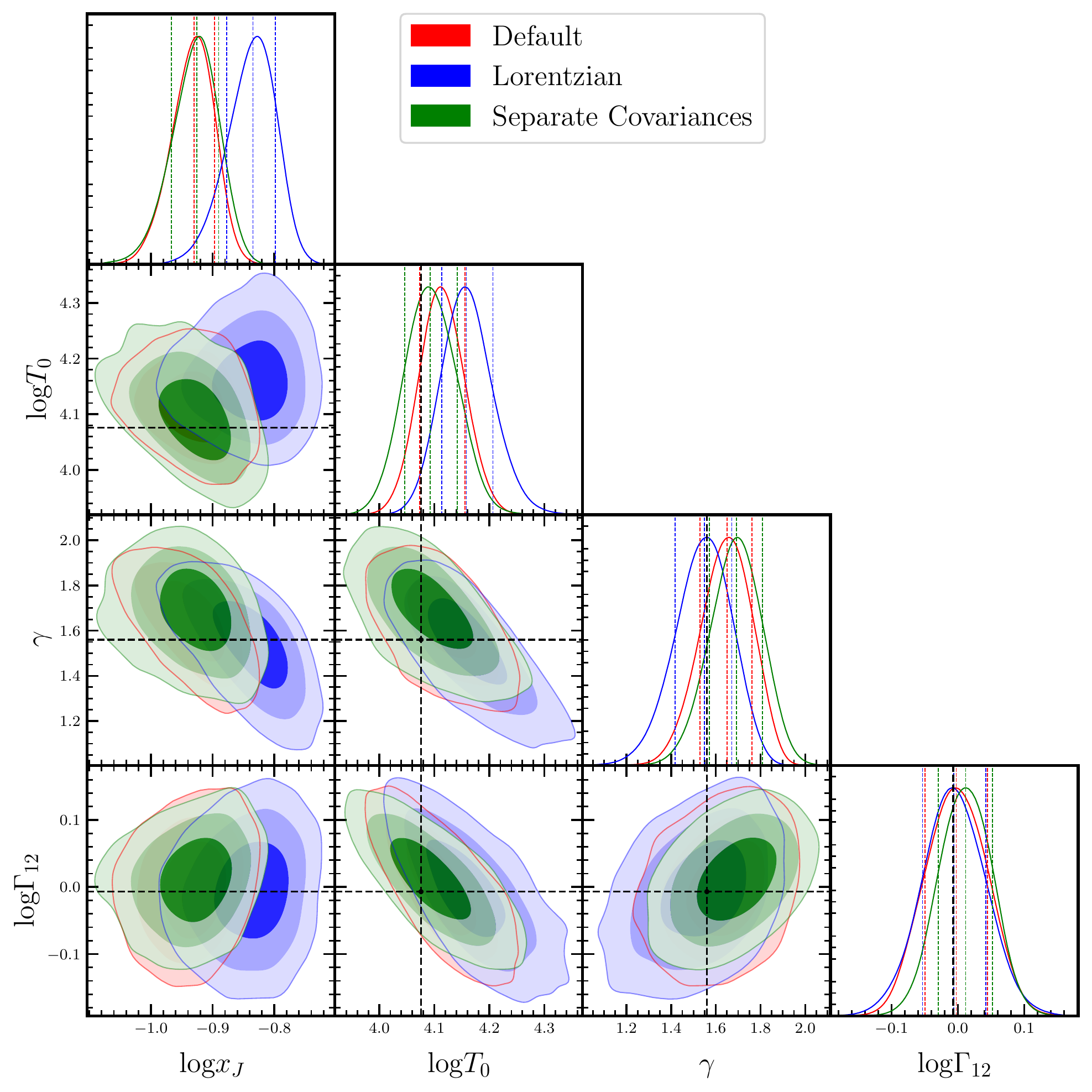}
\caption{Parameter estimates obtained from MCMC runs in a 4D space for default case alongwith Lorentzian smoothing and with separate covariances for FPDF+$\bar{F}$ and FPS.}
\label{fig:corner_cc_gaussian}
\end{figure*}

\begin{figure*}
\centering
\includegraphics[width=1\textwidth]{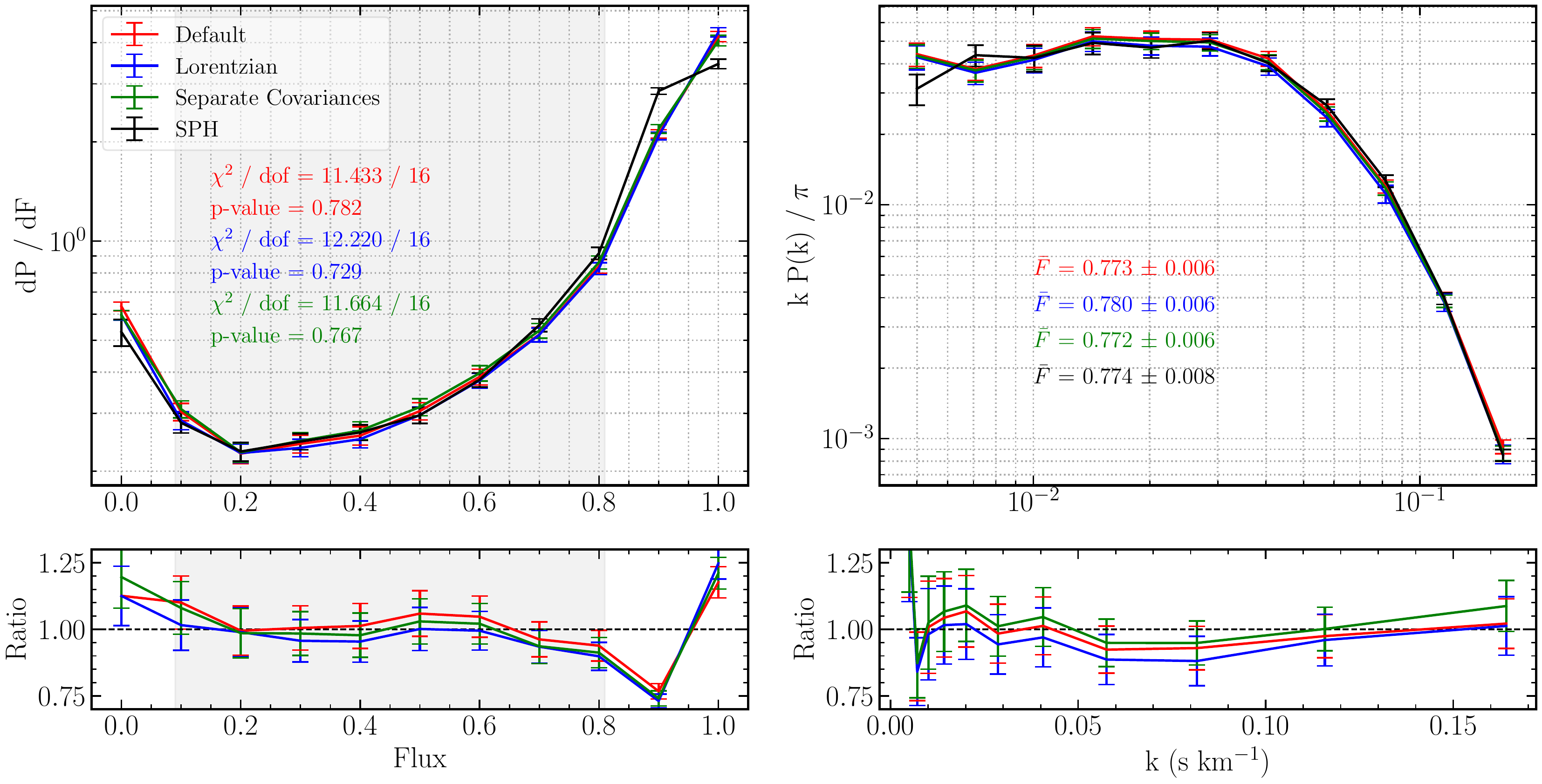}
\caption{Flux statistics for best-fit values of parameters obtained from MCMC run for default alongwith Lorentzian smoothing and with separate covariances for FPDF+$\bar{F}$ and FPS.}
\label{fig:stat_cc_gaussian}
\end{figure*}

In previous section, we had used full covariance matrices for FPDF (plus the mean transmitted flux) and FPS. However, we would also like to explore the effects of ignoring the non-trivial correlations between these two statistics and hence one should take into account two separate covariance matrices, for FPS and \{FPDF+$\bar{F}$\}, each for SPH and lognormal. To do this, let us define the data vector as $X_i = \left\{\mathrm{d}  P(F_i) / \mathrm{d} F, \bar{F} \right\}$ and $P_{\textrm{F}}(k_i)$. The covariance matrices for the sight lines in the SPH simulation can be written as (compare with equations (\ref{eq:cov_sph_jkn}) \& (\ref{eq:cov_sph}))
\begin{multline}
    C^{\mathrm{SPH}, (n)}_{\textrm{FPDF},\bar{F}}(i,j) = \frac{N_\mathrm{spec}-1}{N_\mathrm{spec}} \sum_{k=1}^{N_\mathrm{spec}} \left[\tilde{X}^{(n)}_{-k,i} - \bar{\tilde{X}}^{(n)}_{i}\right] \\ \left[\tilde{X}^{(n)}_{-k,j} - \bar{\tilde{X}}^{(n)}_{j}\right],
    \label{eq:cov_sph_fpdf_jkn}
\end{multline}
\begin{multline}
    C^{\mathrm{SPH}, (n)}_{\textrm{FPS}}(i,j) = \frac{N_\mathrm{spec}-1}{N_\mathrm{spec}} \sum_{k=1}^{N_\mathrm{spec}} \left[\tilde{P}^{(n)}_{\textrm{F},-k,i} - \bar{\tilde{P}}^{(n)}_{\textrm{F},i}\right] \\ \left[\tilde{P}^{(n)}_{\textrm{F},-k,j} - \bar{\tilde{P}}^{(n)}_{\textrm{F},j}\right],
    \label{eq:cov_sph_fps_jkn}
\end{multline}
Following the procedure in eq.\ref{eq:cov_sph}, we can write
\begin{equation}
    C^{\mathrm{SPH}}_{\textrm{FPDF},\bar{F}}(i,j) = \frac{1}{N}\sum_{n=1}^N C^{\mathrm{SPH}, (n)}_{\textrm{FPDF},\bar{F}}(i,j)
     \label{eq:cov_sph_fpdf}
\end{equation}
\begin{equation}
    C^{\mathrm{SPH}}_{\textrm{FPS}}(i,j) = \frac{1}{N}\sum_{n=1}^N C^{\mathrm{SPH}, (n)}_{\textrm{FPS}}(i,j)
     \label{eq:cov_sph_fps}
\end{equation}
Similarly, we can write relations for the covariance matrices of lognormal simulation as well.
\begin{equation}
    C^\mathrm{LN}_{\textrm{FPDF},\bar{F}}(i,j) = \frac{1}{N-1} \sum_{n=1}^{N} \left[\tilde{X}^{(n)}_{i} - \bar{\tilde{X}}_{i}\right] \left[\tilde{X}^{(n)}_{j} - \bar{\tilde{X}}_{j}\right],
    \label{eq:cov_ln_fpdf}
\end{equation}
\begin{equation}
    C^\mathrm{LN}_{\textrm{FPS}}(i,j) = \frac{1}{N-1} \sum_{n=1}^{N} \left[\tilde{P}^{(n)}_{\textrm{F},i} - \bar{\tilde{P}}_{\textrm{F},i}\right] \left[\tilde{P}^{(n)}_{\textrm{F},j} - \bar{\tilde{P}}_{\textrm{F},j}\right],
    \label{eq:cov_ln_fps}
\end{equation}
The total covariance in each case, is given by $C^\mathrm{tot}_{\textrm{FPDF},\bar{F}}(i,j) = C^\mathrm{SPH}_{\textrm{FPDF},\bar{F}}(i,j) + C^\mathrm{SPH}_{\textrm{FPDF},\bar{F}}(i,j)$. An identical relation holds for lognormal simulation as well.
For the likelihood analysis, we compute the $\chi^2$ as
\begin{align}
    \chi^2 &= \left(\mathbf{X}^{\mathrm{LN}} - \mathbf{X}^{\mathrm{SPH}}\right) \left(\mathbf{C}^{\mathrm{tot}}_{\mathrm{FPDF}, \bar{F}} \right)^{-1} \left(\mathbf{X}^{\mathrm{LN}} - \mathbf{X}^{\mathrm{SPH}}\right)^T
    \notag \\
    &+ \left(\mathbf{P}_F^{\mathrm{LN}} - \mathbf{P}_F^{\mathrm{SPH}}\right) \left(\mathbf{C}^{\mathrm{tot}}_{\mathrm{FPS}} \right)^{-1} \left(\mathbf{P}_F^{\mathrm{LN}} - \mathbf{P}_F^{\mathrm{SPH}}\right)^T,
\end{align}
where $\mathbf{\tilde{X}}$ are the row vectors corresponding to the data points $\tilde{X}_i$ and $\mathbf{C}$ are the appropriate covariance matrices. The resulting parameter estimates are shown in Fig. \ref{fig:corner_cc_gaussian} where one compare the results from the default configuration, Gaussian smoothing and full covariance (red contours) with the one with separate covariances (green contours). Fig.\ref{fig:stat_cc_gaussian} shows the statistics for the lognormal model with best-fit parameters and the SPH simulations. From both figures, it is evident that using separate covariances does not make any discernible changes to parameter recovery or fits.

\subsubsection{Lorentzian Smoothing Filter for the baryons}
\noindent

Our default configuration is based on using a Gaussian filter for smoothing the baryonic field, see equation (\ref{eq:Pb_PDM}). However, some people have also used a Lorentzian smoothing filter for reproducing the linear baryonic fluctuations. Hence we carry out a parameter estimation using the Lorentzian smoothing
\begin{equation}
    P_{\mathrm{b}}(k, z) = D^2(z) \frac{P_{\mathrm{DM}}(k)}{[1 + x_{\mathrm{J}}^2(z) k^2]^2},
    \label{eq:Pb_PDM}
\end{equation}
The resulting parameter constraints are shown in Fig. \ref{fig:corner_cc_gaussian} by blue contours. Using a Lorentzian smoothing shifts the constraints on $x_{\textrm{J}}$ to higher values as Lorentzian is a less steeply falling function than Gaussian and hence requires higher values of Jeans length to achieve similar smoothing. The effects of using a different filter are not as pronounced in contours of other three parameters. Fig. \ref{fig:stat_cc_gaussian} shows statistics for best-fit parameters and data. Using a Gaussian smoothing gives a slightly better fit.

\subsubsection{SNR}
\noindent

\begin{figure*}
\centering
\includegraphics[width=0.8\textwidth]{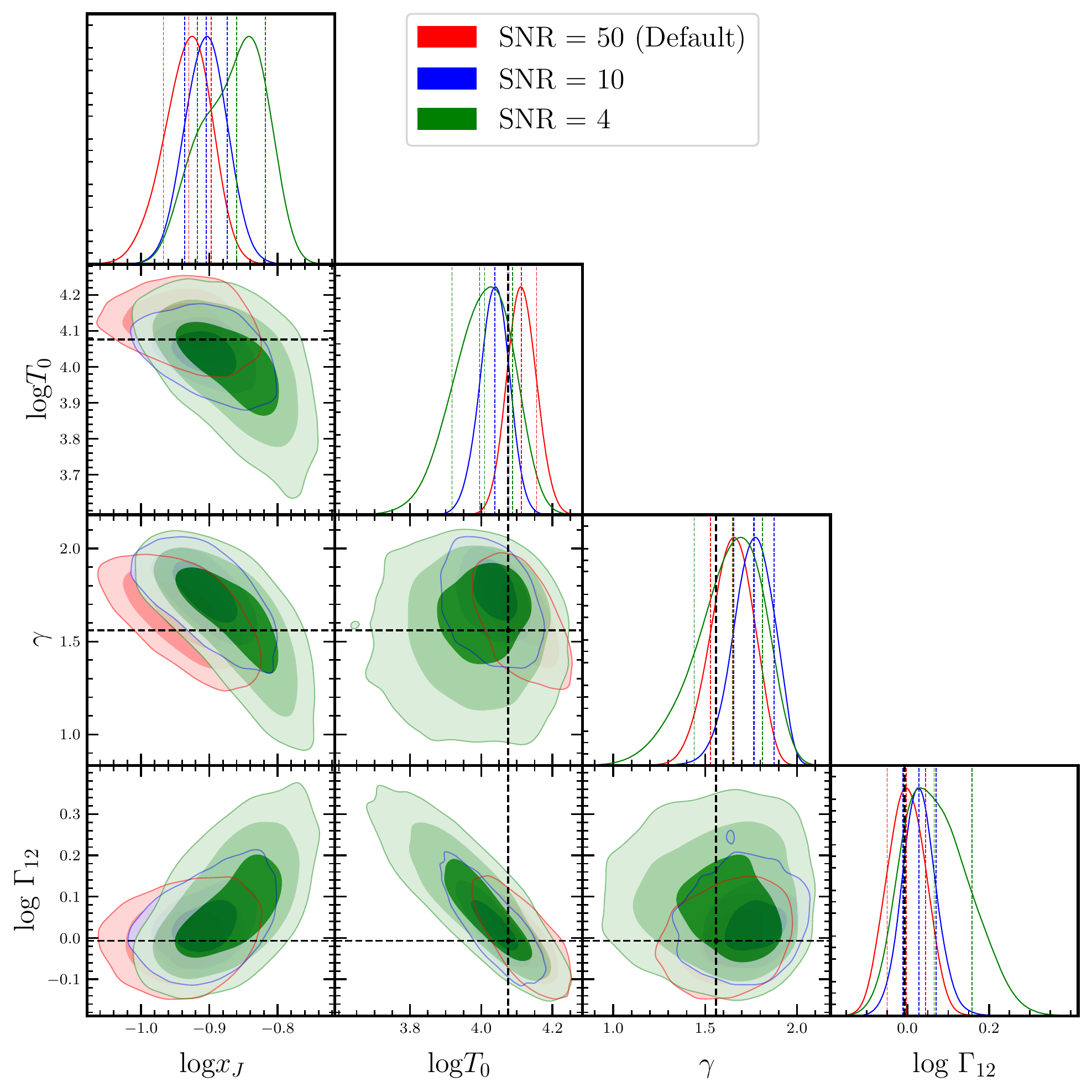}
\caption{Parameter estimates obtained from MCMC runs in a 4D space for three different SNRs.}
\label{fig:corner_SNR}
\end{figure*}

\begin{figure*}
\centering
\includegraphics[width=1\textwidth]{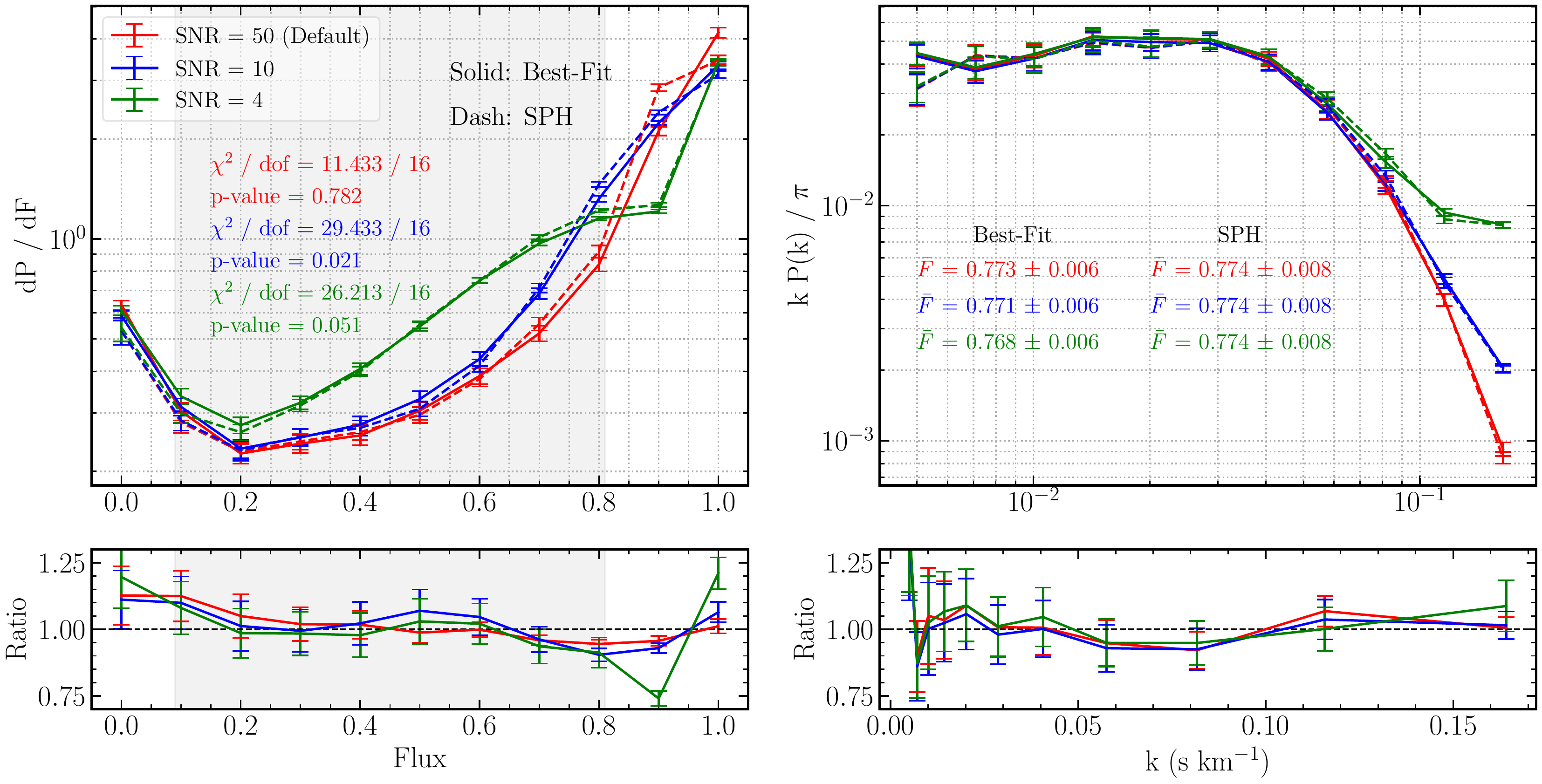}
\caption{Flux statistics for best-fit values of parameters obtained from MCMC run for the three SNRs. Solid (Dashed) lines in top panel show best-fit (SPH) flux statistics.}
\label{fig:stat_SNR}
\end{figure*}

We next discuss the effects of changing SNR on parameter estimates. Fig. \ref{fig:corner_SNR} shows parameter estimates for default value of SNR (=50) along with two different values of SNR (=10 \& 4). Fig. \ref{fig:stat_SNR} shows statistics for best-fit and data for all three SNRs and Table \ref{table:2} summarizes best-fit values for all three SNRs. It is evident from Fig. \ref{fig:corner_SNR} and Table \ref{table:2} that changing SNR does not affect parameter estimates significantly. The right panel of Fig. \ref{fig:stat_SNR} shows the FPS being affected by SNR at small scales with smaller SNRs introducing noise dominated regions at relatively larger scales. The FPS at large scales ($k \lesssim 0.02$ s km$^{-1}$) on the other hand remain almost unaffected by SNR possibly because of cosmic variance dominating over fluctuations due to SNR. The left panel shows that the FPDF in the SPH is strongly affected by changing the SNR; however, the lognormal model correctly tracks these changes.

\begin{table*}
\begin{tabular}{||c c c c c c||} 
 \hline
 Parameter & Prior & True Value & Best-Fit (SNR = 50) & Best-Fit (SNR = 10) & Best-Fit (SNR = 4) \\ [0.5ex]
 \hline\hline
 log $x_{\textrm{J}}$ & [-3.5, 0.5] & - & $-0.921(-0.930)^{+0.033}_{-0.037}$ & $-0.913(-0.905)^{+0.031}_{-0.031}$ & $-0.921(-0.860)^{+0.042}_{-0.057}$ \\ 
 \hline
 log $T_0$ & [2.5, 5.5] & 4.076 & $4.113(4.113)^{+0.042}_{-0.040}$ & $4.054(4.038)^{+0.040}_{-0.042}$ & $4.085(4.009)^{+0.079}_{-0.092}$ \\
 \hline
 $\gamma$ & [-0.5, 5] & 1.56 & $1.616(1.650)^{+0.113}_{-0.121}$ & $1.751(1.767)^{+0.107}_{-0.112}$ & $1.777(1.648)^{+0.164}_{-0.207}$ \\
 \hline
 log $\Gamma_{12}$ & [-2, 2] & -0.007 & $-0.002(-0.002)^{+0.047}_{-0.047}$ & $0.023(0.028)^{+0.042}_{-0.039}$ & $-0.005(0.066)^{+0.092}_{-0.074}$ \\
 \hline
\end{tabular}
\caption{Priors, true and best-fit values for parameter estimates in MCMC runs for three SNRs.}
\label{table:2}
\end{table*}

\subsubsection{SPH with longer path length}
\noindent

\begin{figure*}
\centering
\includegraphics[width=0.8\textwidth]{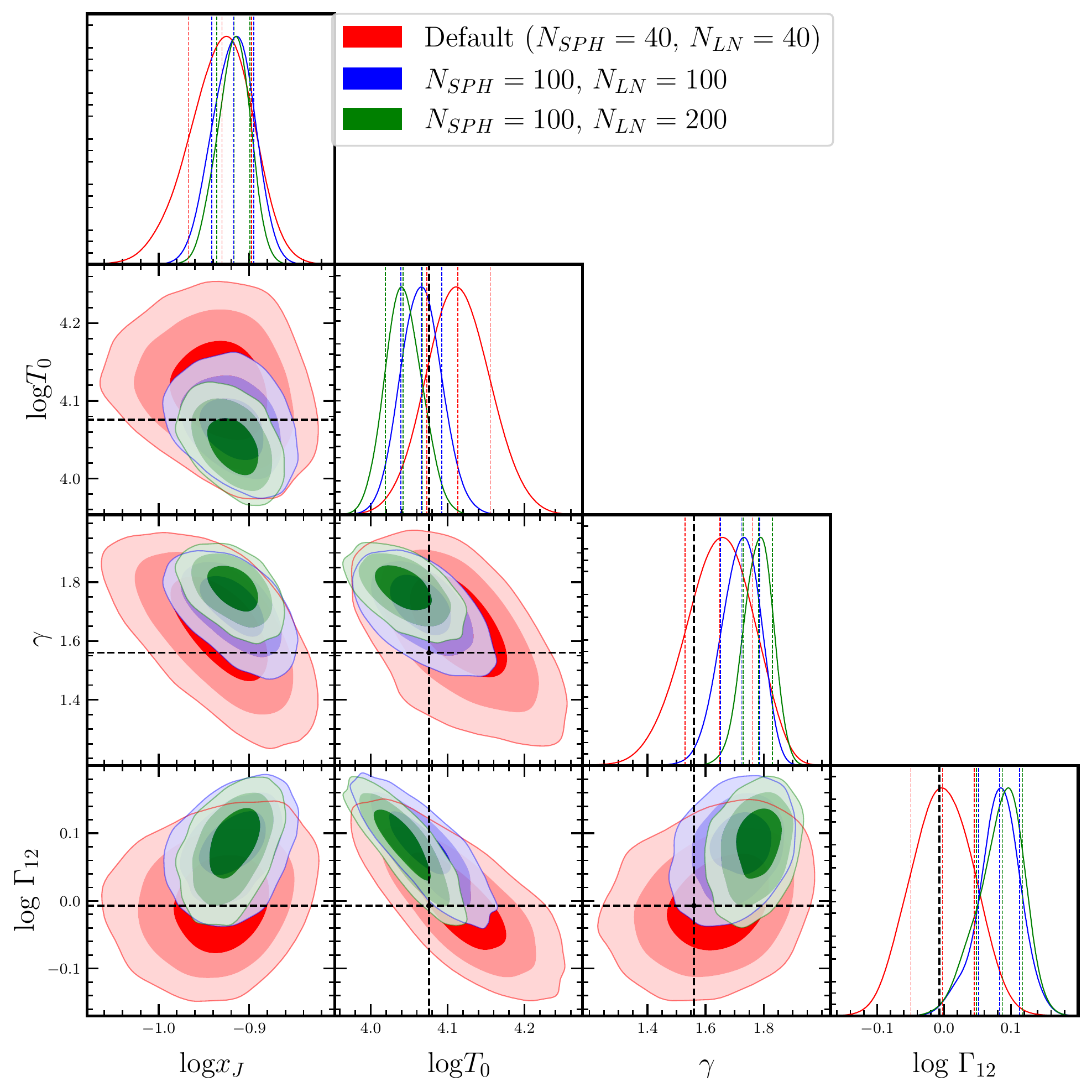}
\caption{Parameter estimates obtained from MCMC runs with SPH having a path length $DX \sim 16.5$ (averaged over 100 sightlines. We use two different path lengths in lognormal, first by averaging over 100 sightlines and secondly, averaging over 200 sightlines. Dotted (solid) black lines show true / inferred (best-fit) values of IGM parameters in simulated data (MCMC run). it is evident that using a larger dataset and reducing uncertainties in lognormal lead to worse fits. Recovery of parameter $T_0$ is still within $1-\sigma$ but $\gamma$ and $\Gamma_{12}$ are now recovered at $\sim 2.5 \sigma$.}
\label{fig:corner_sph100}
\end{figure*}

\begin{figure*}
\centering
\includegraphics[width=1\textwidth]{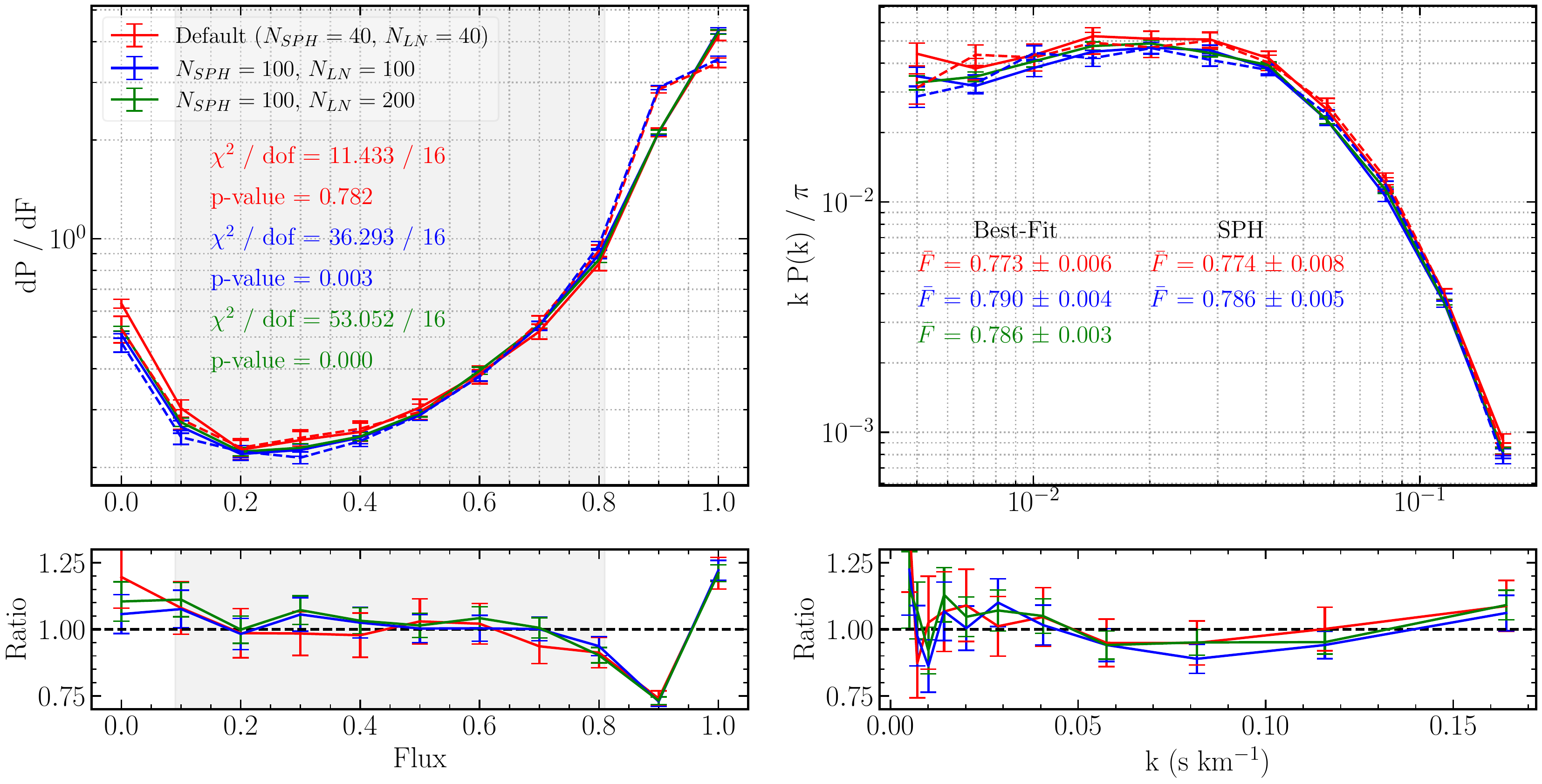}
\caption{Flux statistics for best-fit values of parameters obtained from the MCMC runs corresponding to fig.\ref{fig:corner_sph100}. Solid (Dashed) lines in top panel show best-fit (SPH) flux statistics. A more accurate SPH dataset and reduced uncertainty in lognormal leads to worsened fits, although absolute differences in flux statistics remain similar. Please note that in the case with $N_{\textrm{SPH}} = 100$ and $N_{\textrm{LN}} = 200$ (green curves), the first and last bins in FPDF alone contribute $\sim 34\%$ of total $\chi^2$.}
\label{fig:stat_sph100}
\end{figure*}

In our default configuration, we used a total absorption path length, $DX = 6.2$ (equivalent to averaging over 40 sightlines) in the mock SPH dataset. We then use the same path length for lognormal as well. However, current observational high resolution datasets of the Ly$\alpha$ forest already significantly exceed this value. Therefore, to check the effect of the increased path length on the recovery of parameters, we run two more MCMC chains with total path length $DX = 16.5$ in SPH dataset (equivalent to averaging over $\sim$ 100 sightlines), similar to that used by \citet{goksel_2022}. In the first chain we keep identical path length ($DX = 16.5$) in the lognormal simulations. In the second chain, we try to understand effect of reducing the statistical uncertainties in the lognormal model on parameter recovery and fits. With this in mind, we double the path length in lognormal (which is equivalent to 200 sight lines). The parameter recovery and best-fit flux statistics are shown in Figs.\ref{fig:corner_sph100} and \ref{fig:stat_sph100} respectively. In both the figures, the blue contours and curves represent the first case, while the green ones represent the second case. As we decrease the statistical uncertainties, the recovery of the parameters worsens and also leads to poorer fits to the data. If we concentrate on the case which has the least uncertainty (green contours), we see from Fig.\ref{fig:corner_sph100} that compared to the default configuration (red contours) where all three IGM parameters were recovered within $1-\sigma$, the new estimates for $\gamma$ and $\Gamma_{12}$ are at $\sim 2.5-\sigma$. $T_0$ however, is recovered relatively better at slightly more than $1-\sigma$. In terms of absolute differences, the discrepancy in the recovery of $T_0$ is less than 10\% while that of $\gamma$ and $\Gamma_{12}$ is at $\sim 20\%$.

We have also checked the effect of varying the random seed for selecting the line of sights in the SPH simulations and found that our results remain consistent. We have also checked the effects of changing size of the simulation box (40-1024 \textit{w.r.t.} 80-2048) and resolution (40-2048 \textit{w.r.t.} 40-1024) and found that our results remain consistent.

\section{Conclusions}
\label{sec:conclude}
\noindent

The Ly$\alpha$ forest observed in the spectra of distant QSOs is one of the useful tracers for probing the cosmological matter power spectrum at relatively small scales. The properties of the forest are sensitive to the thermal and ionization state of the IGM and also to the underlying cosmological model. Hence it has been extensively used to constrain parameters related to the astrophysics and cosmology of the IGM. With upcoming large surveys which will allow us to access a large number of QSO spectra, it becomes important to construct theoretical models and simulations which can be used for interpreting the data.

A possible route to construct simulations that are fast and accurate, suitable for parameter exploration, is to replace some of complex physical processes by reasonable assumptions. Such models, based on some assumptions or approximations, need to be validated by comparing with full hydrodynamic simulations so that the parameter estimation carried out using these models are reliable.

In this work, we use such a semi-numerical simulation of the Ly$\alpha$ forest, based on the lognormal model of the density field, and compare the results with a full SPH simulation, namely, the Sherwood simulations. The aim of the work is to verify how well the lognormal model recovers the parameters related to the IGM. In order to do so, we run an MCMC chain using Sherwood simulations as the data for likelihood analysis. The parameter constraints so obtained can be compared with the corresponding values implied in the SPH simulations.

We found that for the default configuration, the recovered values of all three IGM parameters, $T_0$ (temperature corresponding to the mean IGM density), $\gamma$ (the slope of the temperature-density relation) and  $\Gamma_{12}$ (the photoionization rate) are consistent with the values inferred from the SPH simulation to within $1-\sigma$ (9, 4 and 1\% respectively) from median (best-fit) value. Note that $T_0$ in lognormal model does not account for scatter in the equation of state (\ref{eq:T_Delta}).

We also varied some configurations such as separate covariances, baryon smoothing, SNR, data seed and box size and resolution and found that although best-fit and median values of parameters may change significantly, overall correlation structure remains similar.

However, when the size of absorption path length in the SPH data and lognormal model is increased, the recovery of some of the parameters tend to be biased. This analysis indicates the main limitation of the lognormal model in its present form. Although the four IGM parameters can be recovered reasonably well within statistical errors when the path length is relatively smaller, the recovery worsens as the path length is increased to the largest data sets available currently. The results of this paper show that the thermal parameters can be recovered only at the level of $\sim 20\%$ with respect to the input parameters.

In spite of these limitations, the lognormal simulations can have useful applications. Firstly, they provide a rather quick way of inferring the thermal parameters to within $\sim 20\%$, which can then be used for choosing the input parameters in the hydrodynamic simulations. Second, the work opens up the possibility that the lognormal model can be tweaked, e.g., by introducing more parameters, so that the recovery of the IGM parameters become more reliable. This is something we are currently exploring. In addition, we also plan to extend this work to test the validity of the lognormal model for other redshifts and different thermal histories.

\section*{Acknowledgments}
\noindent

We thank R. Srianand for useful discussions in the early phases of this project.
We gratefully acknowledge use of the IUCAA High Performance Computing (HPC) facility.\footnote{\url{http://hpc.iucaa.in}} We thank the Sherwood simulation team for making their data publicly available.

\section*{Data Availability}
The Sherwood simulations are publicly available at \url{https://www.nottingham.ac.uk/astronomy/sherwood/index.php}. The data generated during this work will be made available upon reasonable request to the authors.

\bibliography{references}

\label{lastpage}

\end{document}